\documentclass[a4paper]{spie_margins}

\usepackage{amsmath,amsfonts,amssymb}
\usepackage{graphicx}
\usepackage[colorlinks=true, allcolors=blue]{hyperref}

\usepackage{textcomp}    % used for extra symbols
\usepackage{xcolor}      % used for text colour
\usepackage{aas_macros}  % journal abbreviations (using aas_macros.sty)

\graphicspath{{./images/}}  % store images in a separate folder

\title{The Gravitational-wave Optical Transient Observer (GOTO)}

\author[a]{\mbox{Martin J. Dyer}}
\author[b]{\mbox{Kendall Ackley}}
\author[c]{\mbox{Felipe Jim\'enez-Ibarra}}
\author[b]{\mbox{Joseph Lyman}}
\author[b]{\mbox{Krzysztof Ulaczyk}}
\author[b]{\mbox{Danny Steeghs}}
\author[c]{\mbox{Duncan K. Galloway}}
\author[a,d]{\mbox{Vik S. Dhillon}}
\author[e]{\mbox{Paul O'Brien}}
\author[f]{\mbox{Gavin Ramsay}}
\author[g]{\mbox{Kanthanakorn Noysena}}
\author[h]{\mbox{Rubina Kotak}}
\author[i]{\mbox{Rene Breton}}
\author[j]{\mbox{Laura Nuttall}}
\author[d]{\mbox{Enric Pallé}}
\author[b]{\mbox{Don Pollacco}}
\author[h]{\mbox{Tom Killestein}}
\author[b]{\mbox{Amit Kumar}}
\author[b]{\mbox{David O'Neill}}
\author[j]{\mbox{Lisa Kelsey}}
\author[b]{\mbox{Ben Godson}}
\author[a]{\mbox{Dan Jarvis}}
\author[ ]{\mbox{the GOTO Collaboration}}

\affil[a]{Department of Physics and Astronomy, University of Sheffield, Sheffield S3 7RH, UK}
\affil[b]{Department of Physics, University of Warwick, Coventry CV4 7AL, UK}
\affil[c]{School of Physics \& Astronomy, Monash University, Clayton VIC 3800, Australia}
\affil[d]{Instituto de Astrofísica de Canarias, E-38205 La Laguna, Tenerife, Spain}
\affil[e]{School of Physics \& Astronomy, University of Leicester, University Road, Leicester LE1 7RH, UK}
\affil[f]{Armagh Observatory \& Planetarium, College Hill, Armagh, BT61 9DG, UK}
\affil[g]{National Astronomical Research Institute of Thailand, 260 Moo 4, T. Donkaew, A. Maerim, Chiangmai, 50180
Thailand}
\affil[h]{Department of Physics \& Astronomy, University of Turku, Vesilinnantie 5, Turku, FI-20014, Finland}
\affil[i]{Jodrell Bank Centre for Astrophysics, Department of Physics and Astronomy, The University of Manchester, Manchester M13 9PL, UK}
\affil[j]{Institute of Cosmology \& Gravitation, University of Portsmouth, Portsmouth PO1 3FX, UK}

\authorinfo{Send correspondence to MJD - email: martin.dyer@sheffield.ac.uk}

\begin{document} 
\maketitle

%% Abstract
\begin{abstract}
The Gravitational-wave Optical Transient Observer (GOTO) is a project dedicated to identifying optical counterparts to gravitational-wave detections using a network of dedicated, wide-field telescopes.
After almost a decade of design, construction, and commissioning work, the GOTO network is now fully operational with two antipodal sites: La Palma in the Canary Islands and Siding Spring in Australia. Both sites host two independent robotic mounts, each with a field-of-view of 44 square degrees formed by an array of eight 40~cm telescopes, resulting in an instantaneous 88 square degree field-of-view per site. All four telescopes operate as a single integrated network, with the ultimate aim of surveying the entire sky every 2--3 days and allowing near-24-hour response to transient events within a minute of their detection. 
In the modern era of transient astronomy, automated telescopes like GOTO form a vital link between multi-messenger discovery facilities and in-depth follow-up by larger telescopes. GOTO is already producing a wide range of scientific results, assisted by an efficient discovery pipeline and a successful citizen science project: Kilonova Seekers.
%The Gravitational-wave Optical Transient Observer (GOTO) is a network of wide-field robotic telescopes, designed to detect the optical counterparts of gravitational-wave sources and other multi-messenger events. Each GOTO telescope has a combined field of view of 44 square degrees, formed from an array of eight 40~cm OTAs. The first two mounts (GOTO-1 and GOTO-2) were deployed at the Roque de los Muchachos Observatory on La Palma, Canary Islands, in 2021, and in 2023 two additional mounts (GOTO-3 and GOTO-4) were commissioned at Siding Spring Observatory in New South Wales, Australia. As a complete network, the four telescopes are linked to form a single multi-site observatory. Combined the network can survey the entire visible sky every two nights, and is able to carry out rapid follow-up observations and detections of transient sources.
\end{abstract}

% Keywords 
%\keywords{telescopes -- gravitational waves -- transient follow-up -- sky surveys -- multi-site observatories}
\keywords{robotic telescopes -- sky surveys -- gravitational-wave counterparts -- electromagnetic follow-up -- multi-site observatories -- wide-field telescopes -- telescope arrays -- observatories}

%%%%%%%%%%%%%%%%%%%%%%%%%%%%%%%%%
\section{The GOTO Network}
\label{sec:network}

The Gravitational-wave Optical Transient Observer (GOTO) collaboration\footnote{\url{https://goto-observatory.org/}} was founded in 2014, with the intent of constructing a global network of robotic telescopes to follow-up gravitational-wave detections and hunt for optical counterparts\cite{goto2020,goto2022}. To achieve this, a robotic mount was designed to hold an array of ``unit telescopes'' (UTs): small, off-the-shelf telescopes which could be combined to form a large field of view in a scalable and cost-effective manner.

\newpage

\begin{figure}[t]
    \begin{center}
        \includegraphics[width=0.77\linewidth]{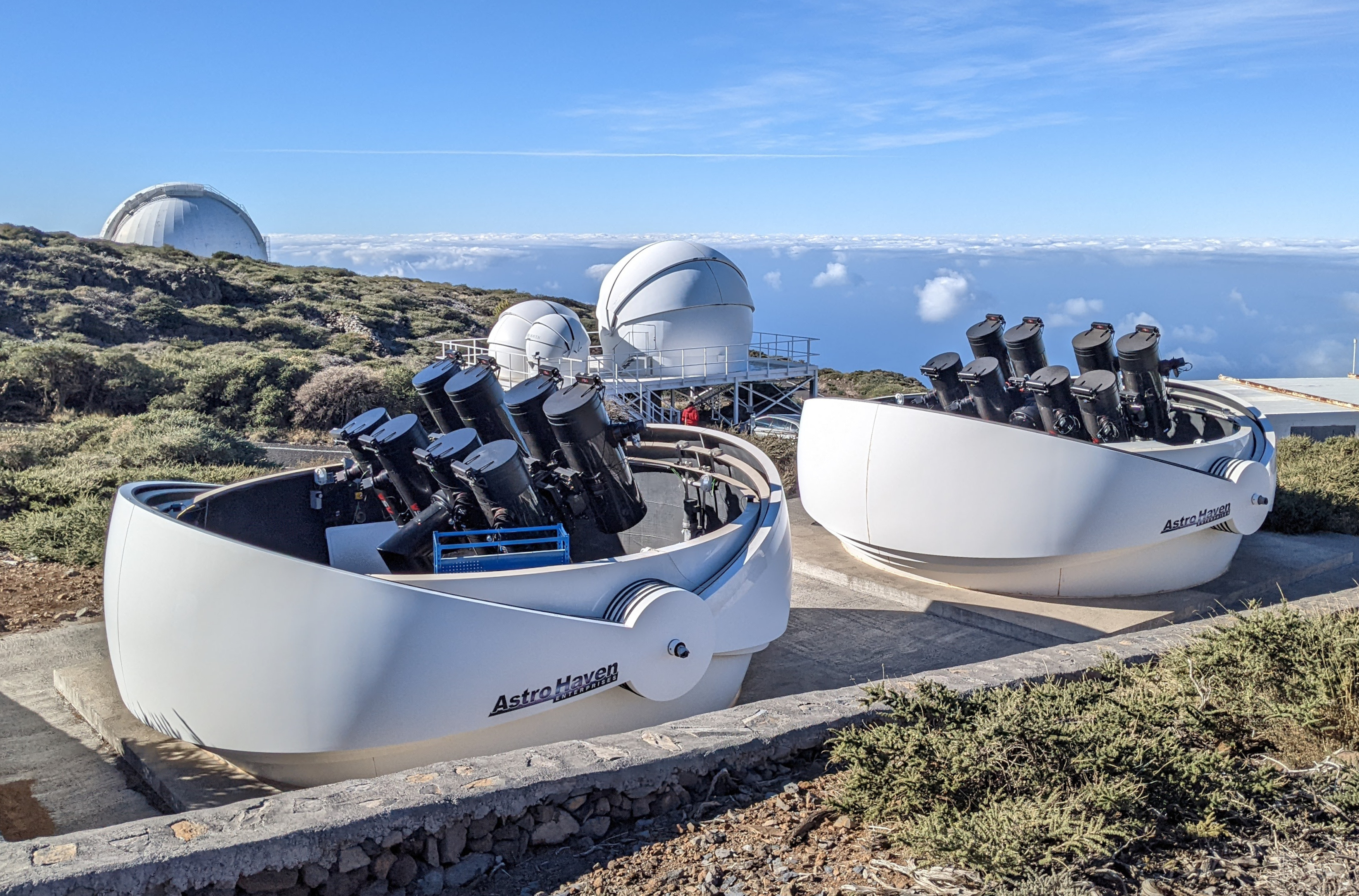}
        \includegraphics[width=0.77\linewidth]{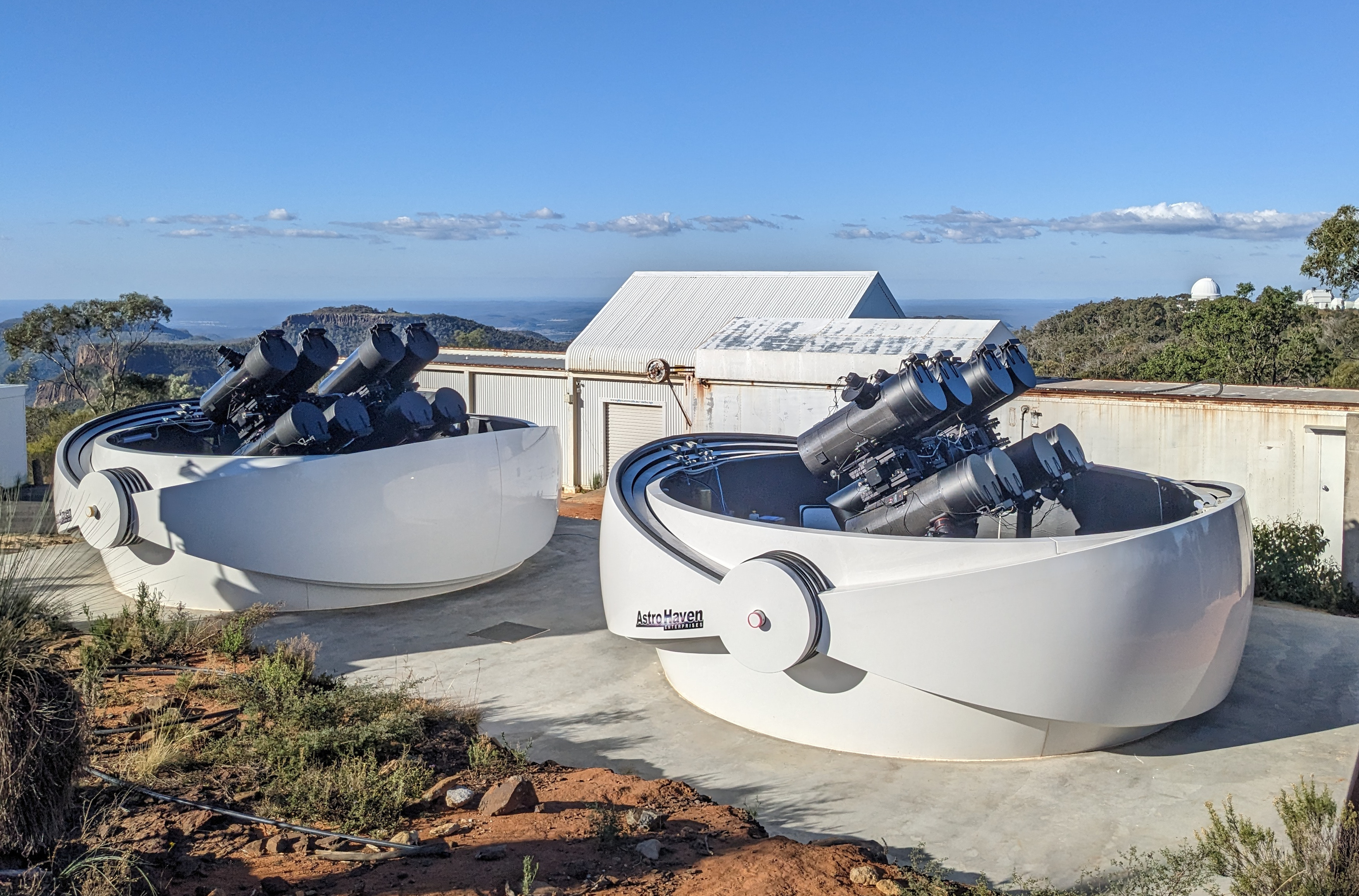}
    \end{center}
    \caption{
        The complete GOTO network as of April 2023. Top: GOTO-North on La Palma, with GOTO-1 on the left and GOTO-2 on the right. Bottom: GOTO-South at Siding Spring, with GOTO-3 on the left and GOTO-4 on the right.
    }\label{fig:photos}
\end{figure}

A prototype telescope with four UTs was constructed at the Roque de los Muchachos Observatory on La Palma, Canary Islands, in 2017\cite{prototype}. Following an extensive commissioning phase, a second-generation telescope with a full array of eight UTs was designed and installed in a neighbouring dome in 2021. Later that year, the prototype was also replaced with a second-generation telescope, forming the complete GOTO-North node. In 2023 two further telescopes were installed at Siding Spring Observatory in New South Wales, Australia, forming the GOTO-South node. These two antipodal sites allow the network to continuously monitor the sky with near 24-hour coverage. All four GOTO telescopes are shown in Figure~\ref{fig:photos}.

\newpage

%%%%%%%%%%%%%%%%%%%%%%%%%%%%%%%%%
\section{Telescope Hardware and Operations}
\label{sec:hardware}

\begin{figure}[t]
    \begin{center}
        \includegraphics[width=0.9\linewidth]{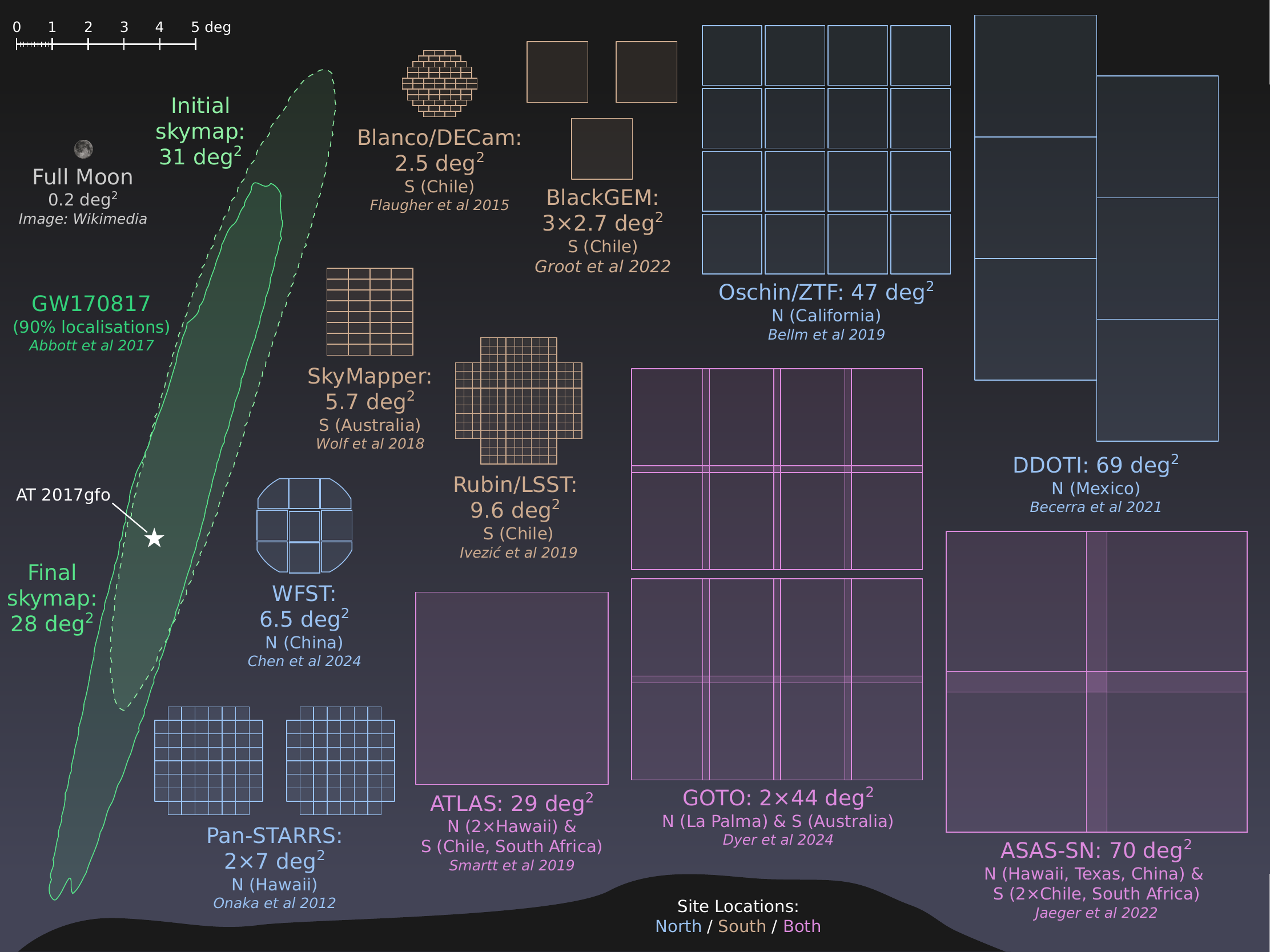}
    \end{center}
    \caption{
        The field of view of selected wide-field telescopes. The depicted footprints represent the sky coverage available from each site; some projects (BlackGEM, GOTO, Pan-STARRS) have multiple mounts per site and some (ASAS-SN, ATLAS, GOTO) have multiple sites across the globe, with the colouring showing if the project has telescopes located in the northern (blue), southern (tan) or both (purple) hemispheres. Included projects: ASAS-SN\cite{ASAS-SN}, ATLAS\cite{ATLAS}, BlackGEM\cite{BlackGEM}, DDOTI\cite{DDOTI}, DECam\cite{DECam}, GOTO (this work), LSST\cite{LSST}, Pan-STARRS\cite{Pan-STARRS}, SkyMapper\cite{SkyMapper}, WFST\cite{WFST} and ZTF\cite{ZTF}. The initial and final skymaps for GW170817\cite{GW170817} (green) and the location of AT~2017gfy (white star) are also shown.
    }\label{fig:fov}
\end{figure}

The second-generation GOTO telescopes were built by ASA Astrosysteme, and are housed in AstroHaven clamshell domes. Each telescope consists of a direct-drive DDM500 German equatorial mount, with a boom-arm holding an array of eight unit telescopes, f2.4, 40~cm H400 Wynne–Riccardi astrographs, as shown in Figure~\ref{fig:photos}. Each UT has a focuser, filter wheel (with Baader $LRGBC$ filters), and an FLI ML50100 camera with a 50 megapixel KAF-50100 CCD detector, and has a field of view of 2.21$^\circ$ $\times$ 2.95$^\circ$. The eight UTs on each mount are aligned to form a tiled array with a small amount of overlap between each field, combined these give a field of view of 44~square~degrees as shown in Figure~\ref{fig:fov}.

GOTO's standard survey observations consist of four 45~s exposures in the wide (400-700 nm) $L$ filter\footnote{The GOTO filter profiles are available on the SVO Filter Profile Service at \url{http://svo2.cab.inta-csic.es/theory/fps/index.php?id=GOTO}.}, when stacked reaching a depth of 20~mag in dark time. With two independent mounts on each site, GOTO can observe an instantaneous area 88~square~degrees. During normal operations, the GOTO telescopes carry out an all-sky survey in the $L$ filter, and together the two antipodal sites cover the entire visible sky every 2--3 nights. This ensures recent comparison images are available to help distinguish potential counterparts during targeted event follow-up searches.

\newpage

%%%%%%%%%%%%%%%%%%%%%%%%%%%%%%%%%
\section{Automation and Analysis Software}
\label{sec:automation}

\begin{figure}[t]
    \begin{center}
        \includegraphics[width=0.95\linewidth]{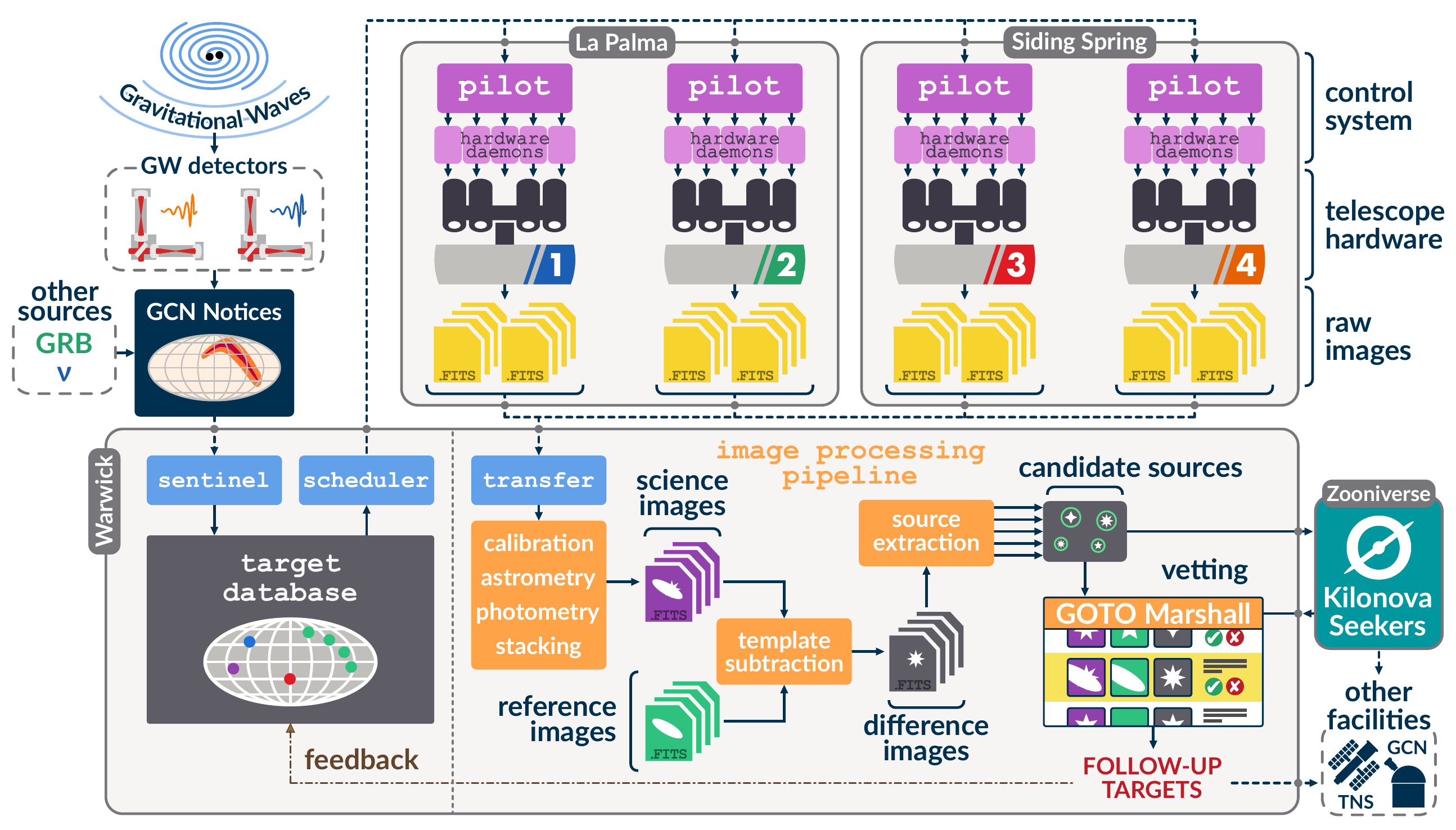}
    \end{center}
    \caption{
        A flow chart visualising the GOTO network. Gravitational wave and other external transient alerts are received and inserted into the target database, which the central scheduler uses to direct observations to the individual telescopes. Raw images are then transferred back to Warwick for processing, with candidate sources being sent to the GOTO Marshall and Kilonova Seekers project for vetting.
    }\label{fig:flow}
\end{figure}

Each GOTO telescope operates autonomously, while target scheduling is coordinated by a central system located at Warwick University in the UK\cite{thesis}. A visual representation of the network is shown in Figure~\ref{fig:flow}. The individual telescopes are operated using the GOTO Telescope Control System (G-TeCS) \cite{gtecs2018,gtecs2020}, with each using a ``pilot'' control program to issue commands and monitor feedback from individual hardware daemons. The pilots each receive pointings from the central scheduler, which ranks and distributes targets to each telescope. Outside the regular all-sky survey, targeted follow-up observations can be triggered by GCN Notices\cite{GCN} received by the ``sentinel'' alert monitor. Alerts produced by the LIGO-Virgo-KAGRA gravitational-wave detector network\cite{LVK} are ranked highest, but other sources include GRB alerts from \textit{Fermi}-GBM\cite{Fermi}, \textit{Swift}-BAT\cite{Swift} and \textit{GECAM}\cite{GECAM}, and neutrino detections from IceCube\cite{IceCube}.  All observations are aligned to a grid of 1048 fixed sky positions, so processed images can be subtracted from archival reference images in order to detect any new sources. 

\begin{figure}[t]
    \begin{center}
        \includegraphics[width=0.89\linewidth]{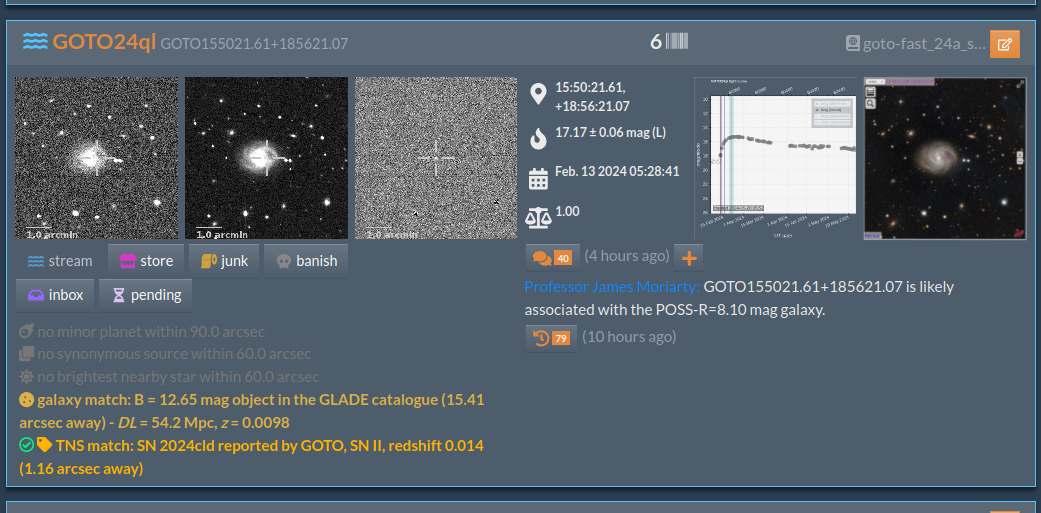}
    \end{center}
    \caption{
        A screenshot of the GOTO Marshall web interface. The source shown is the type II supernova SN2024cld, which was discovered as part of the GOTO-FAST survey\cite{2024cld1,2024cld2}.
    }\label{fig:marshall}
\end{figure}

After source extraction and machine filtering, any candidate sources are sent to the GOTO Marshall web interface for vetting by project members\cite{prototype}. The Marshall, as shown in Figure~\ref{fig:marshall}, provides an easy method for members of the project to vet sources, and includes any historical detections at that point as well as contextual information such as nearby galaxies. Since July 2023 GOTO has also operated the Kilonova Seekers citizen science project\footnote{\url{http://kilonova-seekers.org/}}, with members of the public able to help classify and flag potential candidates. The volunteers are presented with a subset of the information available on the Marshall, including the discovery, reference, and difference images, and are asked to vote on if the source appears to be real. Kilonova Seekers has produced multiple discoveries of real sources that would otherwise have been missed, and the resulting data sets produced are being used to train new automated classifiers, which will further improve the efficiency and discovery speed of the network\cite{KNSeekers}.

\begin{figure}[t]
    \begin{center}
        \includegraphics[width=0.89\linewidth]{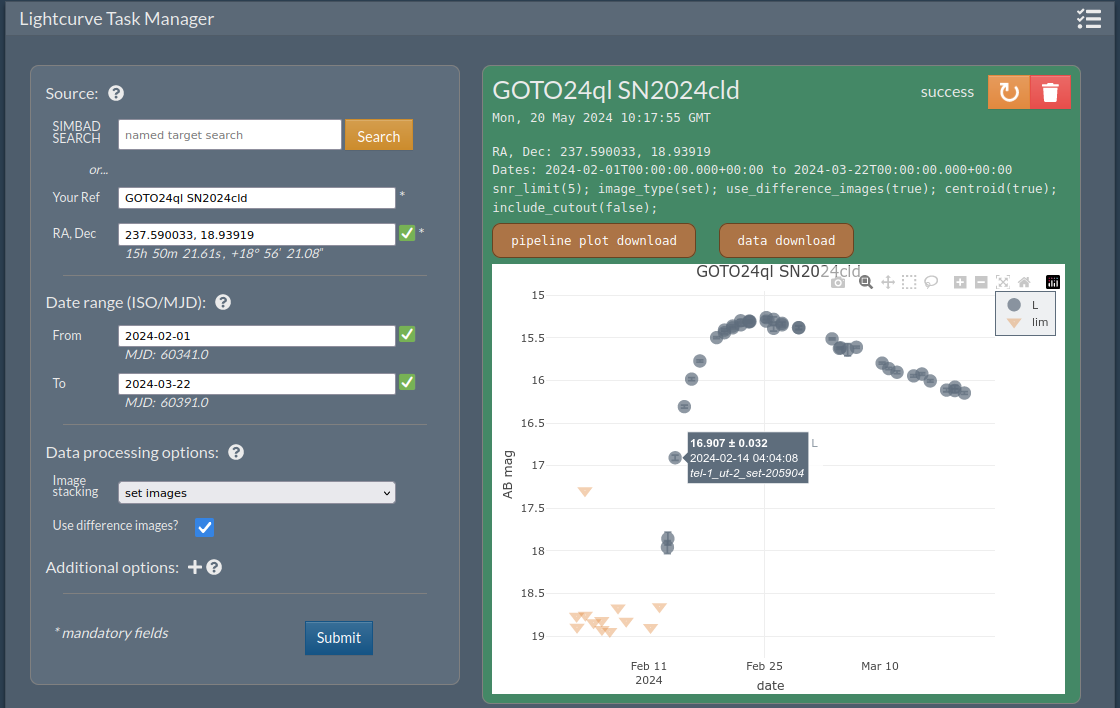}
    \end{center}
    \caption{
        A screenshot of the in-development GOTO Lightcurve web interface for the forced-photometry service.
    }\label{fig:lightcurve}
\end{figure}

Another ongoing project being developed within the GOTO collaboration is a robust forced-photometry service, with the aim of making a publicly accessible API and web interface similar to those produced by ASAS-SN\cite{ASAS-SNforced}, ATLAS\cite{ATLASforced} and ZTF\cite{ZTFforced}. A screenshot of the in-development Lightcurve web interface for forced photometry is shown in Figure~\ref{fig:lightcurve}.

\clearpage

%%%%%%%%%%%%%%%%%%%%%%%%%%%%%%%%%%%%%%%
\section{Results and Future Developments}
\label{sec:conclusion}

We have presented the GOTO project, a new network of wide-field robotic telescopes across the globe dedicated to detecting electromagnetic counterparts to gravitational-wave events and other multi-messenger alerts. The GOTO prototype responded to alerts from the Third LVK Observing Run (O3)\cite{Gompertz2020}, and since 2023 the full network has been following-up alerts from the ongoing O4 run. Other observing campaigns include regular follow-up of gamma-ray bursts\cite{Mong2020,Patel2023}. The fully-automated nature of the network means that, when targets are visible, GOTO is able to start observations within 30~seconds of an alert being received. In April 2024 GOTO observed the candidate gravitational-wave event S240422ed, starting observations within 3 minutes of the trigger and covering 96.2\% of the skymap within the first night\cite{S240422ed}. GOTO-discovered transients are now regularly reported to the Transient Name Server\cite{TNS}, exceeding 100 reports per month in May 2024, and GOTO has already entered the top 10 sources of transient since 2016\footnote{\url{https://www.wis-tns.org/stats-maps}}.

Ongoing efforts to classify GOTO discoveries has also led to the GOTO-FAST program on the Isaac Newton Telescope, which produced immediate spectroscopic classifications of GOTO-discovered sources\cite{GOTO-FAST}. The GOTO Marshall includes quick follow-up triggers for the Liverpool Telescope\cite{Liverpool} and pt5m\cite{pt5m} on La Palma, and other links are being developed with the ANU 2.3m telescope\cite{ANU} at Siding Spring and the future New Robotic Telescope\cite{NRT}.

Now that the GOTO Network is fully operational, the efforts of the collaboration are focused on continuing to follow-up transient alerts, as well as optimising the automated systems and preparing more public utilities including the forced-photometry service.  With the fundamentals in place, GOTO should be well positioned to operate and make discoveries for many years to come.

%%%%%%%%%%%%%%%%%%%%%%%%%%%%%%%%%
\section*{Acknowledgements}

MJD is funded by the UK Science and Technology Facilities Council (STFC; grant number ST/V000853/1). The Gravitational-wave Optical Transient Observer (GOTO) project acknowledges the support of the Monash– Warwick Alliance; the University of Warwick; Monash University; the University of Sheffield; the University of Leicester; Armagh Observatory \& Planetarium; the National Astronomical Research Institute of Thailand (NARIT); the Instituto de Astrofísica de Canarias (IAC); the University of Portsmouth; the University of Turku; the University of Manchester and the UK Science and Technology Facilities Council (grant numbers ST/T007184/1, ST/T003103/1, and ST/T000406/1).

%%%%%%%%%%%%%%%%%%%%%%%%%%%%%%%%%
\bibliography{report}

\begin{thebibliography}{10}

\bibitem{goto2020}
{Dyer}, M.~J., et~al., ``{The Gravitational-wave Optical Transient Observer (GOTO)},'' in [{\em Society of Photo-Optical Instrumentation Engineers (SPIE) Conference Series}{\nolinebreak\hspace{0.1em}]},  {\em Society of Photo-Optical Instrumentation Engineers (SPIE) Conference Series} {\bf 11445},  114457G (Dec. 2020).

\bibitem{goto2022}
{Dyer}, M.~J., et~al., ``{The Gravitational-wave Optical Transient Observer (GOTO)},'' in [{\em Ground-based and Airborne Telescopes IX}{\nolinebreak\hspace{0.1em}]},  {Marshall}, H.~K., et~al., eds., {\em Society of Photo-Optical Instrumentation Engineers (SPIE) Conference Series} {\bf 12182},  121821Y (Aug. 2022).

\bibitem{prototype}
{Steeghs}, D., et~al., ``{The Gravitational-wave Optical Transient Observer (GOTO): prototype performance and prospects for transient science},'' {\em \mnras}~{\bf 511},  2405--2422 (Apr. 2022).

\bibitem{ASAS-SN}
{Shappee}, B.~J., et~al., ``{The Man behind the Curtain: X-Rays Drive the UV through NIR Variability in the 2013 Active Galactic Nucleus Outburst in NGC 2617},'' {\em \apj}~{\bf 788},  48 (June 2014).

\bibitem{ATLAS}
{Tonry}, J.~L., et~al., ``{ATLAS: A High-cadence All-sky Survey System},'' {\em \pasp}~{\bf 130},  064505 (June 2018).

\bibitem{BlackGEM}
{Groot}, P.~J., et~al., ``{BlackGEM: the wide-field multi-band optical telescope array},'' in [{\em Ground-based and Airborne Telescopes IX}{\nolinebreak\hspace{0.1em}]},  {Marshall}, H.~K., et~al., eds., {\em Society of Photo-Optical Instrumentation Engineers (SPIE) Conference Series} {\bf 12182},  121821V (Aug. 2022).

\bibitem{DDOTI}
{Becerra}, R.~L., et~al., ``{DDOTI observations of gravitational-wave sources discovered in O3},'' {\em \mnras}~{\bf 507},  1401--1420 (Oct. 2021).

\bibitem{DECam}
{Flaugher}, B., et~al., ``{The Dark Energy Camera},'' {\em \aj}~{\bf 150},  150 (Nov. 2015).

\bibitem{LSST}
{Ivezi{\'c}}, {\v{Z}}., et~al., ``{LSST: From Science Drivers to Reference Design and Anticipated Data Products},'' {\em \apj}~{\bf 873},  111 (Mar. 2019).

\bibitem{Pan-STARRS}
{Onaka}, P., et~al., ``{GPC1 and GPC2: the Pan-STARRS 1.4 gigapixel mosaic focal plane CCD cameras with an on-sky on-CCD tip-tilt image compensation},'' in [{\em High Energy, Optical, and Infrared Detectors for Astronomy V}{\nolinebreak\hspace{0.1em}]},  {Holland}, A.~D. et~al., eds., {\em Society of Photo-Optical Instrumentation Engineers (SPIE) Conference Series} {\bf 8453},  84530K (July 2012).

\bibitem{SkyMapper}
{Wolf}, C., et~al., ``{SkyMapper Southern Survey: First Data Release (DR1)},'' {\em \pasa}~{\bf 35},  e010 (Feb. 2018).

\bibitem{WFST}
{Chen}, Y.-P., et~al., ``{Basic Survey Scheduling for the Wide Field Survey Telescope (WFST)},'' {\em Research in Astronomy and Astrophysics}~{\bf 24},  015003 (Jan. 2024).

\bibitem{ZTF}
{Bellm}, E.~C., et~al., ``{The Zwicky Transient Facility: System Overview, Performance, and First Results},'' {\em \pasp}~{\bf 131},  018002 (Jan. 2019).

\bibitem{GW170817}
Abbott, B.~P., et~al., ``{GW170817: Observation of Gravitational Waves from a Binary Neutron Star Inspiral},'' {\em Phys. Rev. Lett.}~{\bf 119},  161101 (Oct. 2017).

\bibitem{thesis}
{Dyer}, M.~J., {\em {A telescope control and scheduling system for the Gravitational-wave Optical Transient Observer}}, PhD thesis, University of Sheffield (Jan. 2020).

\bibitem{gtecs2018}
{Dyer}, M.~J., et~al., ``{A telescope control and scheduling system for the Gravitational-wave Optical Transient Observer (GOTO)},'' in [{\em Observatory Operations: Strategies, Processes, and Systems VII}{\nolinebreak\hspace{0.1em}]},  {\em Society of Photo-Optical Instrumentation Engineers (SPIE) Conference Series} {\bf 10704},  107040C (July 2018).

\bibitem{gtecs2020}
{Dyer}, M.~J., et~al., ``{Developing the GOTO telescope control system},'' in [{\em Society of Photo-Optical Instrumentation Engineers (SPIE) Conference Series}{\nolinebreak\hspace{0.1em}]},  {\em Society of Photo-Optical Instrumentation Engineers (SPIE) Conference Series} {\bf 11452},  114521Q (Dec. 2020).

\bibitem{GCN}
{Barthelmy}, S.~D., et~al., ``{The GRB coordinates network (GCN): A status report},'' in [{\em Gamma-Ray Bursts, 4th Hunstville Symposium}{\nolinebreak\hspace{0.1em}]},  {\em American Institute of Physics Conference Series} {\bf 428},  99--103 (May 1998).

\bibitem{LVK}
{Abbott}, B.~P., et~al., ``{Prospects for observing and localizing gravitational-wave transients with Advanced LIGO, Advanced Virgo and KAGRA},'' {\em Living Reviews in Relativity}~{\bf 23},  3 (Sept. 2020).

\bibitem{Fermi}
{Meegan}, C., et~al., ``{The Fermi Gamma-ray Burst Monitor},'' {\em \apj}~{\bf 702},  791--804 (Sept. 2009).

\bibitem{Swift}
{Barthelmy}, S.~D., et~al., ``{The Burst Alert Telescope (BAT) on the SWIFT Midex Mission},'' {\em \ssr}~{\bf 120},  143--164 (Oct. 2005).

\bibitem{GECAM}
Li, X.~Q. et~al., ``{The technology for detection of gamma-ray burst with GECAM satellite},'' {\em Radiat. Detect. Technol. Methods}~{\bf 6}(1),  12--25 (2022).

\bibitem{IceCube}
{Aartsen}, M.~G., et~al., ``{The IceCube Neutrino Observatory: instrumentation and online systems},'' {\em Journal of Instrumentation}~{\bf 12},  P03012 (Mar. 2017).

\bibitem{KNSeekers}
{Killestein}, T.~L. et~al., ``{Kilonova Seekers: the GOTO project for real-time citizen science in time-domain astrophysics},'' (in prep.).

\bibitem{2024cld1}
{Pursiainen}, M., et~al., ``{GOTO Transient Discovery Report for 2024-02-13},'' {\em Transient Name Server Discovery Report}~{\bf 2024-422},  1 (Feb. 2024).

\bibitem{2024cld2}
{Godson}, B., et~al., ``{GOTO Transient Classification Report for 2024-02-13},'' {\em Transient Name Server Classification Report}~{\bf 2024-429},  1 (Feb. 2024).

\bibitem{ASAS-SNforced}
{Kochanek}, C.~S., et~al., ``{The All-Sky Automated Survey for Supernovae (ASAS-SN) Light Curve Server v1.0},'' {\em \pasp}~{\bf 129},  104502 (Oct. 2017).

\bibitem{ATLASforced}
{Shingles}, L., et~al., ``{Release of the ATLAS Forced Photometry server for public use},'' {\em Transient Name Server AstroNote}~{\bf 7},  1--7 (Jan. 2021).

\bibitem{ZTFforced}
{Masci}, F.~J., et~al., ``{A New Forced Photometry Service for the Zwicky Transient Facility},'' {\em arXiv e-prints} ,  arXiv:2305.16279 (May 2023).

\bibitem{Gompertz2020}
{Gompertz}, B.~P., et~al., ``{Searching for electromagnetic counterparts to gravitational-wave merger events with the prototype Gravitational-Wave Optical Transient Observer (GOTO-4)},'' {\em \mnras}~{\bf 497},  726--738 (July 2020).

\bibitem{Mong2020}
{Mong}, Y.~L., et~al., ``{Machine learning for transient recognition in difference imaging with minimum sampling effort},'' {\em \mnras}~{\bf 499},  6009--6017 (Oct. 2020).

\bibitem{Patel2023}
{Patel}, M., et~al., ``{GRB 201015A and the nature of low-luminosity soft gamma-ray bursts},'' {\em \mnras}~{\bf 523},  4923--4937 (Aug. 2023).

\bibitem{S240422ed}
{Ackley}, K., et~al., ``{LIGO/Virgo/KAGRA S240422ed: GOTO counterpart search},'' {\em GRB Coordinates Network}~{\bf 36257},  1 (Apr. 2024).

\bibitem{TNS}
{Gal-Yam}, A., ``{The TNS alert system},'' in [{\em American Astronomical Society Meeting Abstracts}{\nolinebreak\hspace{0.1em}]},  {\em American Astronomical Society Meeting Abstracts} {\bf 237},  423.05 (Jan. 2021).

\bibitem{GOTO-FAST}
{Godson}, B. et~al., ``{GOTO-FAST: Fast Analysis and Spectroscopy of Transients},'' (in prep.).

\bibitem{Liverpool}
{Steele}, I.~A., et~al., ``{The Liverpool Telescope: performance and first results},'' in [{\em Ground-based Telescopes}{\nolinebreak\hspace{0.1em}]},  {Oschmann}, Jacobus~M., J., ed., {\em Society of Photo-Optical Instrumentation Engineers (SPIE) Conference Series} {\bf 5489},  679--692 (Oct. 2004).

\bibitem{pt5m}
{Hardy}, L.~K., et~al., ``{pt5m - a 0.5 m robotic telescope on La Palma},'' {\em \mnras}~{\bf 454},  4316--4325 (Dec. 2015).

\bibitem{ANU}
{Dopita}, M., et~al., ``{The Wide Field Spectrograph (WiFeS): performance and data reduction},'' {\em \apss}~{\bf 327},  245--257 (June 2010).

\bibitem{NRT}
{Jermak}, H.~E., et~al., ``{The New Robotic Telescope: progress report},'' in [{\em Ground-based and Airborne Telescopes VIII}{\nolinebreak\hspace{0.1em}]},  {Marshall}, H.~K., et~al., eds., {\em Society of Photo-Optical Instrumentation Engineers (SPIE) Conference Series} {\bf 11445},  114453D (Dec. 2020).

\end{thebibliography}
\bibliographystyle{spiebib_oneauthor}

%%%%%%%%%%%%%%%%%%%%%%%%%%%%%%%%%

\end{document}